\documentclass[aps,pra,notitlepage,superscriptaddress,showpacs]{revtex4-2}
\usepackage[T1]{fontenc}
\usepackage{amsmath}
\usepackage{amssymb}
\usepackage{graphicx}
\usepackage{esint}
\begin{document}
\title{Adiabatons in double tripod coherent atom-light coupling scheme }
\author{Via\v{c}eslav Kudria\v{s}ov}
\email{viaceslav.kudriasov@ff.vu.lt}

\affiliation{Institute of Theoretical Physics and Astronomy, Vilnius University,
Saul\.etekio 3, Vilnius LT-10257, Lithuania}
\author{Hamid R. Hamedi}
\email{hamid.hamedi@tfai.vu.lt}

\affiliation{Institute of Theoretical Physics and Astronomy, Vilnius University,
Saul\.etekio 3, Vilnius LT-10257, Lithuania}
\author{Julius Ruseckas}
\email{julius.ruseckas@bpti.eu }

\affiliation{Baltic Institute of Advanced Technology, Pilies St. 16-8, LT-01403
Vilnius, Lithuania}
\begin{abstract}
Optical adiabatons are specific shape-invariant pulse pairs propagating
at the reduced group velocity and without optical absorption in the
medium. The purpose of this study is to analyze and demonstrate adiabaton
formation in many level atomic systems. Here we focus on the analysis
of five level M-type and double tripod systems. It is found that M-type
atomic systems are prone to intensity dependent group velocity and
pulse front steepening which prevents the formation of long range
optical adiabatons. In contrast, the double tripod atomic system is
quite favorable for the formation of optical adiabatons leading to
two different optical field configurations propagating with invariant
shape.
\end{abstract}
\maketitle

\section{Introduction}

Precise control of light-matter interactions is crucial in optics
research. Remarkable progress has been made using quantum interference
in resonant media \citep{Li2001,2005-RMP-Fleischhauer,Dmitry2017},
where materials can be manipulated into specific quantum states to
enhance or suppress light absorption. One notable technique that exemplifies
this approach is Electromagnetically Induced Transparency (EIT) \citep{1997-PT-Harris,2005-RMP-Fleischhauer,Wu2005EIT,Hamedi2017}.
EIT allows a material to become transparent to a specific light frequency
that it would normally absorb. This technique underscores the power
of quantum interference in achieving precise control over the optical
response of materials.

Beyond transparency, EIT can lead to a wealth of fascinating effects,
including the formation of nonabsorbing dark states \citep{1996-PO-Arimondo},
slow group velocities (slow light) \citep{1999-Nature-Hau}, giant
enhancement of nonlinearities \citep{Wang2001,2003-RMP-Lukin,Niu2006,Hamedi2015}
and other phenomena \citep{Paspalakis2002OL,Paspalakis2002,Kis2003,Wu2004,Ying2004,Li2006,Clader2007,Wu2007,Chen2014}.
Most of these effects can be easily observed in a simple three- or
four-level atomic medium where two or three laser fields (probe and
pump) coherently interact with the atomic system tuned to resonant
optical transitions. More sophisticated interaction effects with richer
dynamics can be observed in light-atom configurations operating with
a larger number of atomic levels. This extends the possibilities of
coherent control significantly. 

Typically, EIT-based research focuses on conditions where the probe
field is much weaker than the coupling field. However, in practical
applications, such as achieving extremely low group velocities, the
intensity of the coupling laser must be kept quite low. This requirement
presents a challenge for ultra-slow light propagation because the
probe field needs to be significantly weaker than the coupling field
to maintain the validity of the weak probe approximation. Consequently,
the probe field also needs to be strong enough to be detectable, complicating
the situation as the weak probe approximation is often not met.

This issue has led to an interest in studying the effects of stronger
probe fields in EIT scenarios. Grobe et al. \citep{Grobe1994} addressed
this by finding an exact solution to the coupled nonlinear Maxwell-Schr\"odinger
equations for the $\Lambda$ scheme under adiabatic conditions. This
solution permits waves of arbitrary shapes to propagate, retaining
their shapes relatively unchanged after some initial reshaping. These
shape-preserving waves are known as adiabatons. An optical adiabaton
is a pair of shape-invariant pulses with slowly changing envelopes
that propagate at reduced group velocities without optical absorption
in the medium. Adiabatons represent a nonlinear pulse propagation
regime under EIT conditions, where the intensity of the probe field
is comparable to that of the control field, allowing the adiabatic
approximation to hold \citep{Fleischhauer1996}. When the system deviates
from adiabaticity, the pulses experience absorption, attenuation,
reshaping, and steepening effects. 

The theoretical prediction of adiabatons in a $\Lambda$ system \citep{Grobe1994}
was soon experimentally confirmed \citep{1995-PRL-Kasapi}. Subsequent
studies explored this phenomenon in various systems and conditions
\citep{Cerboneschi1996,2004-PRA-Buffa,2006-PRA-Arkhipkin,2006-PRA-Shakhmuratov,2005-PRA-Mazets,Hioe2008}.
For instance, a four-level double-$\Lambda$ scheme was used to demonstrate
pulse matching and adiabaton-type propagation \citep{Cerboneschi1996}.
Temporal reshaping and compression of a probe pulse, due to its group
velocity's dependence on the control field, were also studied \citep{2004-PRA-Buffa,2006-PRA-Arkhipkin}.
Two types of adiabatons with zero and non-zero energy in $\Lambda$
and ladder schemes, respectively, were analyzed \citep{2006-PRA-Shakhmuratov}.
Additionally, two pairs of adiabatically propagating pulses at different
group speeds\textemdash fast and slow adiabatons\textemdash were found
in a tripod system \citep{2005-PRA-Mazets}. Adiabatons for five-level
systems have only been analyzed in a simple five-level double-$\Lambda$
scheme \citep{Hioe2008}. scheme. This motivates us to explore the
formation of adiabatons in more complicated five-level schemes with
more laser fields involved. The presence of additional control fields
will enhance the controllability in the formation and propagation
of shape-preserving adiabatons.

In this paper, we theoretically investigate the optical pulse propagation
under the adiabatic regime in other known multilevel atomic systems:
M-type \citep{LumingM,Ying2006M} and double-tripod \citep{2010-PRL-Unanyan,Ruseckas2011,Ruseckas2013,Bao2011,2014-NC-Lee}
. Both M-type and double tripod systems have five distinct energy
levels, but different number of light fields arranged in an M-shape
or double tripod interaction configuration, respectively. The latter
scheme has some remarkable features compared to a simple three level
$\Lambda$ system. Particularly, in a double tripod system there are
two probe beams at different frequencies which couple together via
four coupling fields and atomic coherences. This property leads to
the formation of a two-component or spinor slow light (SSL) having
some distinct properties \citep{2010-PRL-Unanyan,2014-NC-Lee}. In
this kind of medium only the special combinations of the probe fields
(normal modes) propagate in the atomic cloud with the definite (and
different) velocities. Our study shows that the double tripod configuration
supports the adiabatic pulse propagation by exhibiting two different
configurations of the fields that propagate without changing their
shape. This regime, however, is not supported in all multi-level systems
that have an uncoupled (dark) state, for example, it is not possible
in the M-type coupling scheme.

The paper is structured as follows. To illustrate our approach, we
first derive some well-established analytical results and the adiabatic
approximation, as previously described in Refs.~\citep{Grobe1994,Fleischhauer1996},
for a simple $\Lambda$ system containing only two optical fields
(Sec.~\ref{sec:lambda}). This serves as a basis for our later analysis
of more complex multilevel systems. Next, we proceed to theoretical
analysis of systems with more energy levels and additional coupled
fields. In Sec.~\ref{sec:M-type} we examine light pulse propagation
in a multilevel M-type system, which notably does not exhibit adiabaton
formation. Then, in Sec.~\ref{sec:Double-tripod} we present calculations
demonstrating adiabatonic propagation in a double tripod system. Finally,
we conclude with a discussion and summarize our findings in Sec.~\ref{sec:Conclusions}. 

\section{Adiabatons in $\Lambda$-type atomic system \label{sec:lambda}}

\subsection{Equations of motion for atoms and fields}

\begin{figure}
\includegraphics[width=0.5\columnwidth]{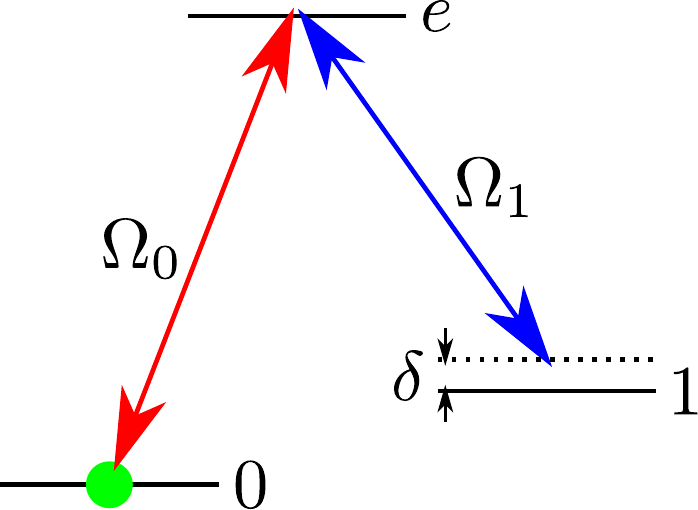} \caption{Three level $\Lambda$-type atomic system. Two laser beams with the
Rabi frequencies $\Omega_{0}$ and $\Omega_{1}$ act on atoms characterized
by two hyperfine ground levels $0$ and $1$ as well as an excited
level $e$. Parameter $\delta$ denotes two-photon detuning from resonance.
Atoms are initially in the ground level $0$.}
\label{lambda}
\end{figure}

We consider a $\Lambda$-type atomic system, shown in Fig.~\ref{lambda},
involving two metastable ground states $|0\rangle$ and $|1\rangle$,
as well as an excited state $|e\rangle$. Laser fields with the Rabi
frequncies $\Omega_{0}$ and $\Omega_{1}$ induce resonant transitions
$|0\rangle\rightarrow|e\rangle$ and $|1\rangle\rightarrow|e\rangle$,
respectively. Applying the rotating wave approximation (RWA), the
atomic Hamiltonian in the rotating frame with respect to the atomic
levels reads

\begin{equation}
H_{\Lambda}=-\frac{1}{2}\left(\Omega_{0}|e\rangle\langle0|+\Omega_{1}|e\rangle\langle1|+\mathrm{H.c.}\right)+\delta|1\rangle\langle1|-\frac{i}{2}\Gamma|e\rangle\langle e|,\label{eq:H}
\end{equation}
where $\delta$ is a two-photon detuning. The losses in the Hamiltinian
(\ref{eq:H}) are taken into account in an effective way by introducing
a rate $\Gamma$ of the excited state decay. In order to simplify
the mathematical description of the system while keeping the relevant
physical details, we characterize the state of an atom using a state
vector $|\Psi\rangle=\psi_{0}|0\rangle+\psi_{1}|1\rangle+\psi_{e}|e\rangle$,
as in Ref.~\citep{Grobe1994,Eberly1995}, instead of more complete
description employed in Refs.~\citep{Fleischhauer1996,Chuang2017}
that involves a density matrix.

The time-dependent Schr\"odinger equation $i\hbar\partial_{t}|\Psi\rangle=H_{\Lambda}|\Psi\rangle$
for the atomic state-vector $|\Psi\rangle$ yields the following equations
for the atomic probability amplitudes $\psi_{0}$, $\psi_{1}$ and
$\psi_{e}$:

\begin{align}
i\partial_{t}\psi_{0} & =-\frac{1}{2}\Omega_{0}^{*}\psi_{e}\,,\label{eq:psi-1}\\
i\partial_{t}\psi_{1} & =\delta\psi_{1}-\frac{1}{2}\Omega_{1}^{*}\psi_{e}\,,\label{eq:psi-2}\\
i\partial_{t}\psi_{e} & =-\frac{i}{2}\Gamma\psi_{e}-\frac{1}{2}\Omega_{0}\psi_{0}-\frac{1}{2}\Omega_{1}\psi_{1}.\label{eq:psi-3}
\end{align}

On the other hand, the Rabi frequencies of the laser fields obey the
propagation equations

\begin{equation}
\partial_{t}\Omega_{l}+c\partial_{z}\Omega_{l}=\frac{i}{2}g\psi_{e}\psi_{l}^{*}\,,\qquad l=0,1\;\label{eq:el-1}
\end{equation}
where the parameter $g$ characterizes the strength of coupling of
the light fields with the atoms. It is related to the optical depth
$\alpha$ as $g=c\Gamma\alpha/L$, where $L$ is the length of the
medium. For simplicity here we have assumed that the coupling strength
$g$ is the same for both laser fields. 

\subsection{Coupled and uncoupled states}

The Hamiltonian (\ref{eq:H}) can be rewitten as \citep{Fleischhauer1996}:

\begin{equation}
H_{\Lambda}=-\frac{\Omega}{2}\left(|e\rangle\langle\mathrm{C}|+|\mathrm{C}\rangle\langle e|\right)+\delta|1\rangle\langle1|-\frac{i}{2}\Gamma|e\rangle\langle e|\,,\label{eq:6}
\end{equation}
where

\begin{equation}
|\mathrm{C}\rangle=\frac{1}{\Omega}(\Omega_{0}^{*}|0\rangle+\Omega_{1}^{*}|1\rangle),\label{eq:C-state}
\end{equation}
is a coupled state with 
\begin{equation}
\Omega=\sqrt{|\Omega_{0}|^{2}+|\Omega_{1}|^{2},}\label{eq:7}
\end{equation}
 being a total Rabi frequency. Additionally we introduce an uncoupled
state

\begin{equation}
|\mathrm{U}\rangle=\frac{1}{\Omega}(\Omega_{1}|0\rangle-\Omega_{0}|1\rangle),\label{eq:U-state}
\end{equation}
which is orthogonal to the coupled state. The probability amplitudes
to find an atom in the coupled and uncoupled states read 

\begin{align}
\psi_{\mathrm{C}} & =\langle\mathrm{C}|\Psi\rangle=\frac{1}{\Omega}(\Omega_{0}\psi_{0}+\Omega_{1}\psi_{1})\,,\label{eq:C-definition}\\
\psi_{\mathrm{U}} & =\langle\mathrm{U}|\Psi\rangle=\frac{1}{\Omega}(\Omega_{1}^{*}\psi_{0}-\Omega_{0}^{*}\psi_{1})\,.\label{eq:U-def}
\end{align}

In terms of coupled and uncoupled states, the equations (\ref{eq:psi-1})-(\ref{eq:psi-3})
for the atomic amplitudes take the form

\begin{align}
i\partial_{t}\psi_{\mathrm{U}} & =\Delta\psi_{\mathrm{U}}+\Omega_{-}^{*}\psi_{\mathrm{C}}\,,\label{eq:U}\\
i\partial_{t}\psi_{\mathrm{C}} & =-\Delta\psi_{\mathrm{C}}+\Omega_{-}\psi_{\mathrm{U}}-\frac{1}{2}\Omega\psi_{e}\,,\label{eq:C}\\
i\partial_{t}\psi_{e} & =-\frac{i}{2}\Gamma\psi_{e}-\frac{1}{2}\Omega\psi_{\mathrm{C}}\,,\label{eq:3-C}
\end{align}
with \citep{Fleischhauer1996}

\begin{align}
\Delta & =i\frac{\Omega_{0}}{\Omega}\partial_{t}\frac{\Omega_{0}^{*}}{\Omega}+i\frac{\Omega_{1}}{\Omega}\partial_{t}\frac{\Omega_{1}^{*}}{\Omega}+\delta\frac{|\Omega_{0}|^{2}}{\Omega^{2}}\,,\label{eq:Delta}\\
\Omega_{-} & =i\frac{\Omega_{1}}{\Omega}\partial_{t}\frac{\Omega_{0}}{\Omega}-i\frac{\Omega_{0}}{\Omega}\partial_{t}\frac{\Omega_{1}}{\Omega}-\delta\frac{\Omega_{0}}{\Omega}\frac{\Omega_{1}}{\Omega}\,.\label{eq:Omega_minus}
\end{align}

Here $\Omega_{-}$ describes non-adiabatic losses and $2\Delta$ represents
the separaton in energies between the uncoupled and coupled states.

\subsection{Adiabatic approximation}

Let us consider a situation where the total Rabi frequency $\Omega$
is sufficiently large so that the conditions presented below by Eqs.~(\ref{eq:adiabat-cond-1})
and (\ref{eq:adiabat-cond-2}) hold. In this case the adiabatic approximation
can be applied. To prepare for the derivation of approximate equations
we express $\psi_{e}$ from Eq.~(\ref{eq:C}):

\begin{equation}
\psi_{e}=2\frac{\Omega_{-}}{\Omega}\psi_{\mathrm{U}}-\frac{2}{\Omega}(i\partial_{t}+\Delta)\psi_{\mathrm{C}}\,.\label{eq:3-CU}
\end{equation}

On the other hand, Eq.~(\ref{eq:3-C}) relates $\psi_{\mathrm{C}}$
to $\psi_{e}$ as: 
\begin{equation}
\psi_{\mathrm{C}}=-\frac{2i}{\Omega}\left(\partial_{t}+\frac{\Gamma}{2}\right)\psi_{e}\,.\label{eq:C-3}
\end{equation}
 Since the excited state decay rate $\Gamma$ is considered to be
large compared to the rate of change of the fields, we neglect the
temporal derivative $\partial_{t}$ in the above equation. Substituting
Eq.~(\ref{eq:3-CU}) into Eq.~(\ref{eq:C-3}) one gets

\begin{equation}
\psi_{\mathrm{C}}=-2i\frac{\Gamma}{\Omega^{2}}(\Omega_{-}\psi_{\mathrm{U}}-(i\partial_{t}+\Delta)\psi_{\mathrm{C}})\,.\label{eq:C-2}
\end{equation}

We solve this equation iteratively with respect to $\psi_{\mathrm{C}}$,
assuming that $\Omega$ is large compared to the rate of non-adiabatic
transitions 

\begin{equation}
\Omega\gg|\Omega_{-}|\,.\label{eq:adiabat-cond-1}
\end{equation}

In the zeroth-order of the adiabatic approximation the coupled state
is not populated, $\psi_{\mathrm{C}}\approx0$, so Eq.~(\ref{eq:C-definition})
yields

\begin{equation}
\psi_{0}\approx\frac{\Omega_{1}}{\Omega}\,,\quad\psi_{1}\approx-\frac{\Omega_{0}}{\Omega}\,.\label{eq:21}
\end{equation}

Non-zero $\psi_{C}$ appears in the first-order approximation. Putting
$\psi_{\mathrm{C}}\approx0$ on the right hand side (r.h.s.) of Eq.~(\ref{eq:C-2}),
one arrives at the first order result for the amplitude of the coupled
state

\begin{equation}
\psi_{\mathrm{C}}\approx-2i\frac{\Gamma}{\Omega}\frac{\Omega_{-}}{\Omega}\psi_{\mathrm{U}}\,.\label{eq:C-U}
\end{equation}

Inserting this expression back in the r.h.s. of Eq.~(\ref{eq:C-2})
we obtain the expression containing the second-order correction

\begin{equation}
\psi_{\mathrm{C}}=-2i\frac{\Gamma}{\Omega}\frac{\Omega_{-}}{\Omega}\psi_{\mathrm{U}}+4\frac{\Gamma^{2}}{\Omega^{2}}(i\partial_{t}+\Delta)\frac{\Omega_{-}}{\Omega^{2}}\psi_{\mathrm{U}}\,.\label{eq:23}
\end{equation}

The second order term should be much smaller than the first order
one in the r.h.s. of the above equation. Since the atomic population
is concentrated in the uncoupled state, $|\psi_{\mathrm{U}}|\approx1$
and $|\psi_{\mathrm{C}}|\ll1$, one arrives at the following condition

\begin{equation}
\frac{\Gamma|\Delta|}{\Omega^{2}}\ll1\,,\qquad\frac{\Gamma|\Omega_{-}|}{\Omega^{2}}\ll1\,.\label{eq:adiabat-cond-2}
\end{equation}

Now let us present the adiabatic expansion of the excited state amplitude
$\psi_{3}$. The zeroth-order approximation of Eq.~(\ref{eq:3-CU})
is $\psi_{e}\approx0$. In the first order, taking into account Eq.~(\ref{eq:C-U})
and the condition (\ref{eq:adiabat-cond-2}), we have

\begin{equation}
\psi_{e}\approx2\frac{\Omega_{-}}{\Omega}\psi_{\mathrm{U}}\,.\label{eq:3-U}
\end{equation}

Since the excited state should be weakly populated, the condition
(\ref{eq:adiabat-cond-1}) is to be imposed.

Finally, inserting the first-order adiabatic result Eq.~(\ref{eq:C-U})
relating $\psi_{\mathrm{C}}$ to $\psi_{\mathrm{U}}$ into Eq.~(\ref{eq:U})
we obtain the equation for the amplitude of the uncoupled state

\begin{equation}
i\partial_{t}\psi_{\mathrm{U}}=\Delta\psi_{\mathrm{U}}-2i\Gamma\frac{|\Omega_{-}|^{2}}{\Omega^{2}}\psi_{\mathrm{U}}\,.\label{eq:main-U}
\end{equation}

The last term on the r.h.s. represents losses due to non-adiabatic
corrections. Similarly, inserting Eqs.~(\ref{eq:3-U}) and (\ref{eq:C-U})
into Eq.~(\ref{eq:el-1}) we get the equations for the amplitudes
of the radiation fields $\Omega_{0}$ and $\Omega_{1}$:

\begin{align}
\partial_{t}\Omega_{0}+c\partial_{z}\Omega_{0} & =g\left(i\frac{\Omega_{-}}{\Omega}\frac{\Omega_{1}^{*}}{\Omega}-2\frac{\Gamma}{\Omega}\frac{|\Omega_{-}|^{2}}{\Omega^{2}}\frac{\Omega_{0}}{\Omega}\right)|\psi_{\mathrm{U}}|^{2}\,,\,\label{eq:main-1}\\
\partial_{t}\Omega_{1}+c\partial_{z}\Omega_{1} & =g\left(-i\frac{\Omega_{-}}{\Omega}\frac{\Omega_{0}^{*}}{\Omega}-2\frac{\Gamma}{\Omega}\frac{|\Omega_{-}|^{2}}{\Omega^{2}}\frac{\Omega_{1}}{\Omega}\right)|\psi_{\mathrm{U}}|^{2}.\label{eq:main-2}
\end{align}

Equations (\ref{eq:main-U}), (\ref{eq:main-1}) and (\ref{eq:main-2})
describe adiabatic propagation of the fields. The second terms on
the r.h.s. of Eqs.~(\ref{eq:main-1}) and (\ref{eq:main-2}) describe
non-adiabatic losses.

\subsection{Short duration of propagation}

Let us now consider the case when the duration of the propagation
$\tau_{\mathrm{prop}}$ is much smaller than the life time of the
adiabatons: $\Gamma\frac{|\Omega_{-}|^{2}}{\Omega^{2}}\tau_{\mathrm{prop}}\ll1$.
The propagation duration is of the order of $L/v_{g}$, where $v_{g}=c\Omega^{2}/g$
is the group velocity. Expressing the atom-light coupling strength
via the optical density $\alpha$, $g=c\Gamma\alpha/L$, we obtain
the condition 

\begin{equation}
\left(\frac{\Gamma|\Omega_{-}|}{\Omega^{2}}\right)^{2}\alpha\ll1,\label{eq:29}
\end{equation}
which should be satisfied. In this situation one can neglect the decay
terms in Eqs.~(\ref{eq:main-U}), (\ref{eq:main-1}) and (\ref{eq:main-2}).
Furthermore, since $|\psi_{\mathrm{U}}|\approx1$, Eqs.~(\ref{eq:main-1}),
(\ref{eq:main-2}) reduce to

\begin{align}
\partial_{t}\Omega_{0}+c\partial_{z}\Omega_{0} & =ig\frac{\Omega_{-}}{\Omega}\frac{\Omega_{1}^{*}}{\Omega}\,,\label{eq:main-1-a}\\
\partial_{t}\Omega_{1}+c\partial_{z}\Omega_{1} & =-ig\frac{\Omega_{-}}{\Omega}\frac{\Omega_{0}^{*}}{\Omega}\,.\label{eq:main-2-a}
\end{align}

Combining Eqs.~(\ref{eq:main-1-a}) and (\ref{eq:main-2-a}), the
total Rabi frequency $\Omega$ obeys the equation

\begin{equation}
\partial_{t}\Omega+c\partial_{z}\Omega=0\,.\label{eq:Omega-const}
\end{equation}

On the other hand, Eqs.~(\ref{eq:main-1-a}) and (\ref{eq:main-2-a})
provide the following equation for the ratio $\chi=\Omega_{0}/\Omega_{1}$:

\begin{equation}
\left(c^{-1}+\frac{g}{c\Omega^{2}}\right)\partial_{t}\chi+\partial_{z}\chi+i\delta\frac{g}{c\Omega^{2}}\chi=0\,.\label{eq:ratio}
\end{equation}

This equation has a similar form to the equation for the propagation
of a weak probe field affected by a stronger control field when $|\Omega_{0}/\Omega_{1}|\ll1$.
Yet, in the present situation, such a condition is not imposed.

\subsection{Solution}

Equation~(\ref{eq:Omega-const}) has the following solution satisfying
the boundary condition at $z=0$:

\begin{equation}
\Omega(z,t)\equiv\Omega(\tau)=\sqrt{|\Omega_{0}(0,\tau)|^{2}+|\Omega_{1}(0,\tau)|^{2}}\,,\label{eq:34}
\end{equation}
where $\tau=t-z/c$. To obtain a solution of Eq.~(\ref{eq:ratio})
we change the variables $t$ and $z$ to $\tau=t-z/c$ and $z$, respectively.
Then

\begin{equation}
\frac{g}{c\Omega^{2}(\tau)}\partial_{\tau}\chi+\partial_{z}\chi+i\delta\frac{g}{c\Omega^{2}(\tau)}\chi=0\,.\label{eq:35}
\end{equation}
Further we change the time variable $\tau$ to the stretched time

\begin{equation}
\zeta(\tau)=\frac{c}{g}\int_{-\infty}^{\tau}\Omega^{2}(\tau')d\tau'\,,\label{eq:stretched-time}
\end{equation}
yielding

\begin{equation}
\partial_{\zeta}\chi+\partial_{z}\chi+i\delta\frac{g}{c\Omega^{2}(\zeta)}\chi=0\,.\label{eq:F-stretched}
\end{equation}

A particular solution of this equation is

\begin{equation}
\chi(z,\zeta)=\exp\left(ik(z-\zeta)-i\delta\frac{g}{c}\int_{-\infty}^{\zeta}\frac{d\zeta'}{\Omega^{2}(\zeta')}\right)\,.\label{eq:38}
\end{equation}

When $\delta=0$, a solution of Eq.~(\ref{eq:F-stretched}) is an
arbitrary function $f(\zeta-z)$, fixed by the boundary condition
at $z=0$:

\begin{equation}
f(\zeta(t))=\frac{\Omega_{0}(0,t)}{\Omega_{1}(0,t)},\label{eq:39}
\end{equation}
or

\begin{equation}
f(z)=\frac{\Omega_{0}(0,\zeta^{-1}(z))}{\Omega_{1}(0,\zeta^{-1}(z))}\,,\label{eq:40}
\end{equation}
where $\zeta^{-1}$ is a function inverse to the function $\zeta$.
It follows from Eq.~(\ref{eq:stretched-time}) that if $\Omega(\tau)$
becomes constant after a certain time, the stretched time $\zeta$
becomes a linear function of $\tau$. After that time the obtained
solution $f(\zeta-z)$ describes a shape-preserving propagation \citep{Grobe1994}.

\begin{figure}
\includegraphics[width=0.7\columnwidth]{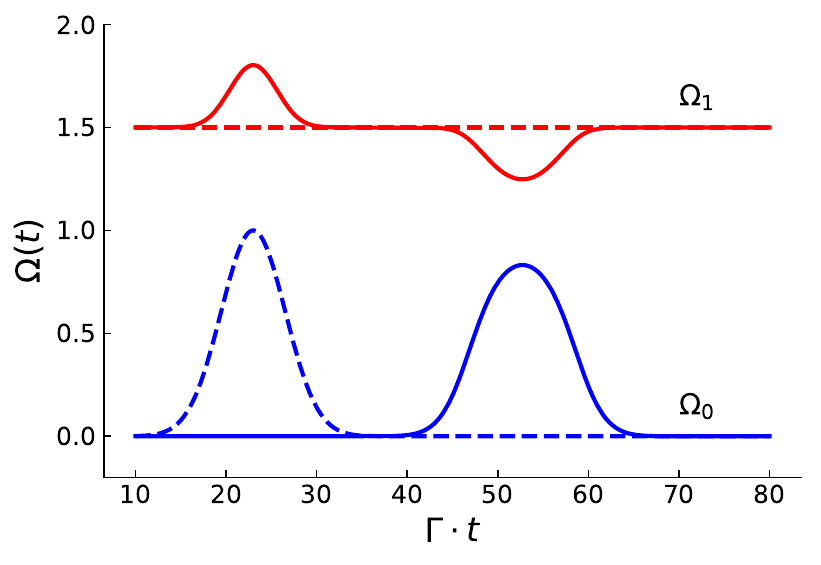}

\caption{Temporal dependence of pulse amplitudes in $\Lambda$ system. Dashed
lines correspond to pulse envelopes $\Omega_{0}$ and $\Omega_{1}$
at the input ($z=0$), solid lines show pulse envelopes at propagation
distance $z=70L_{\mathrm{abs}}$. Here $L_{\mathrm{abs}}=L/\alpha$
is resonant absorption length. Amplitude shown on the vertical scale
is measured in $\Gamma$.}
\label{fig:Lambda-adiabaton}
\end{figure}

We performed numerical investigation of the above analytical study
by modeling field equations with optical fields of particular temporal
shape. Here, the first optical field is described as a Gaussian pulse
at the input: $\Omega_{0}(0,t)=A\exp[-(t-t_{0})^{2}/\tau_{0}^{2}]$,
where $\tau_{0}=5\Gamma^{-1}$, $t_{0}=23\Gamma^{-1}$ and $A=\Gamma$.
The second one is simply a constant field: $\Omega_{1}(0,t)=1.5\Gamma$.
Results of the numerical solution are depicted in Fig.~\ref{fig:Lambda-adiabaton}
which shows the adiabatonic propagation regime for the optical fields
after some propagation time. Specifically, a shape-preserving combination
of fields $\Omega_{0}(z,t)$ and $\Omega_{1}(z,t)$ propagating with
the group velocity $v_{g}$ can be seen on the right part of Fig.~\ref{fig:Lambda-adiabaton}.
The upward pulse from $\Omega_{1}(z,t)$ field seen in the left upper
part of the figure is propagating with the speed of light.

\section{Propagation of light pulses in M-type system \label{sec:M-type}}

\subsection{Equations of motion for atoms and fields}

\begin{figure}
\includegraphics[width=0.5\columnwidth]{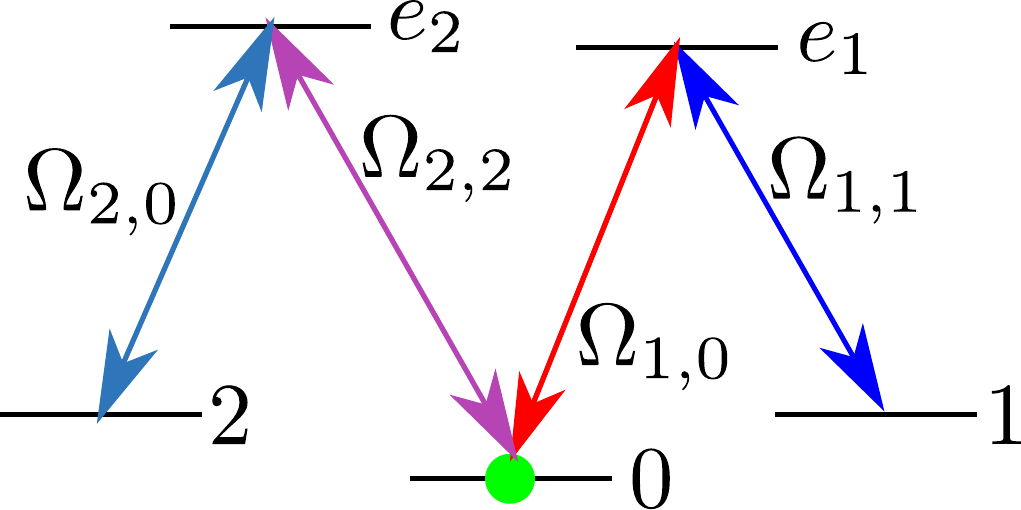}

\caption{Five level M-type atomic system. Four laser beams with the Rabi frequencies
$\Omega_{1,0}$, $\Omega_{1,1}$, $\Omega_{2,2}$ and $\Omega_{2,0}$
act on atoms characterized by three hyperfine ground levels $0$,
$1$ and $2$ as well as two excited levels $e_{1}$ and $e_{2}$.
Atoms are initially in the ground level $0$.}
\label{fig:m-type}
\end{figure}

Now we will consider a M-type atomic system (Fig.~\ref{fig:m-type})
involving three metastable ground states $|0\rangle$, $|1\rangle$
and $|2\rangle$, as well as two excited states $|e_{1}\rangle$ and
$|e_{2}\rangle$. Four laser fields with the Rabi frequencies $\Omega_{j,l}$
induce resonant transitions $|l\rangle\rightarrow|e_{j}\rangle$;
here the $j=1,2$ and $l=0,j$. Applying the rotating wave approximation
(RWA), the atomic Hamiltonian in the rotating frame with respect to
the atomic levels reads

\begin{equation}
H_{\mathrm{M}}=\sum_{j=1}^{2}\left\{ -\frac{1}{2}\left(\Omega_{j,0}|e_{j}\rangle\langle0|+\Omega_{j,j}|e_{j}\rangle\langle j|+\mathrm{H.c.}\right)-\frac{i}{2}\Gamma|e_{j}\rangle\langle e_{j}|+\delta_{j}|j\rangle\langle j|\right\} \,,\label{eq:H-1}
\end{equation}
where $\delta_{j}$ are two-photon detunings. The losses in the Hamiltonian
(\ref{eq:H-1}) are taken into account in an effective way by introducing
a rate $\Gamma$ of the excited state decay. The time-dependent Schr\"odinger
equation $i\hbar\partial_{t}|\Psi\rangle=H_{\mathrm{M}}|\Psi\rangle$
for the atomic state-vector $|\Psi\rangle=\sum_{l=0}^{2}\psi_{l}|l\rangle+\sum_{j=1}^{2}\psi_{e_{j}}|e_{j}\rangle$
yields the following equations for the atomic probability amplitudes
$\psi_{l}$, $\psi_{e_{j}}$:

\begin{align}
i\partial_{t}\psi_{0} & =-\frac{1}{2}\Omega_{1,0}^{*}\psi_{e_{1}}-\frac{1}{2}\Omega_{2,0}^{*}\psi_{e_{2}}\,,\label{eq:42}\\
i\partial_{t}\psi_{1} & =\delta_{1}\psi_{1}-\frac{1}{2}\Omega_{1,1}^{*}\psi_{e_{1}}\,,\label{eq:43}\\
i\partial_{t}\psi_{2} & =\delta_{2}\psi_{2}-\frac{1}{2}\Omega_{2,2}^{*}\psi_{e_{2}}\,,\label{eq:44}\\
i\partial_{t}\psi_{e_{1}} & =-\frac{i}{2}\Gamma\psi_{e_{1}}-\frac{1}{2}\Omega_{1,0}\psi_{0}-\frac{1}{2}\Omega_{1,1}\psi_{1}\,,\label{eq:45}\\
i\partial_{t}\psi_{e_{2}} & =-\frac{i}{2}\Gamma\psi_{e_{2}}-\frac{1}{2}\Omega_{2,0}\psi_{0}-\frac{1}{2}\Omega_{2,2}\psi_{2}\,.\label{eq:46}
\end{align}

On the other hand, the Rabi frequencies of the laser fields obey the
following propagation equations

\begin{align}
(\partial_{t}+c\partial_{z})\Omega_{1,0} & =\frac{i}{2}g\psi_{e_{1}}\psi_{0}^{*}\,,\label{eq:el10}\\
(\partial_{t}+c\partial_{z})\Omega_{2,0} & =\frac{i}{2}g\psi_{e_{2}}\psi_{0}^{*}\,,\label{eq:48}\\
(\partial_{t}+c\partial_{z})\Omega_{1,1} & =\frac{i}{2}g\psi_{e_{1}}\psi_{1}^{*}\,,\label{eq:49}\\
(\partial_{t}+c\partial_{z})\Omega_{2,2} & =\frac{i}{2}g\psi_{e_{2}}\psi_{2}^{*}\,,\label{eq:el22}
\end{align}
where the parameter $g$ characterizes the strength of coupling of
the light fields with the atoms.

\subsection{Coupled and uncoupled states}

The Hamiltonian (\ref{eq:H-1}) can be rewitten as

\begin{equation}
H_{\mathrm{M}}=\sum_{j=1}^{2}\left\{ -\frac{1}{2}\left(\Omega_{j,j}|e_{j}\rangle\langle\mathrm{C}_{j}|+\Omega_{j,j}^{*}|\mathrm{C}_{j}\rangle\langle e_{j}|\right)+\delta_{j}|j\rangle\langle j|-\frac{i}{2}\Gamma|e_{j}\rangle\langle e_{j}|\right\} \,,\label{eq:H-M}
\end{equation}
where $|\mathrm{C}_{j}\rangle$ are two coupled states which are unnormalized
and non-orthogonal to each other

\begin{equation}
|\mathrm{C}_{j}\rangle=\chi_{j}^{*}|0\rangle+|j\rangle\,,\quad\mathrm{with\quad\chi_{j}=\frac{\Omega_{j,0}}{\Omega_{j,j}},}\quad j=1,2\,.\label{eq:C-state-1}
\end{equation}

Additionally we introduce a normalized uncoupled state which is orthogonal
to both coupled states

\begin{equation}
|\mathrm{U}\rangle=\frac{1}{N_{0}}(|0\rangle-\chi_{1}|1\rangle-\chi_{2}|2\rangle)\,,\label{eq:U-state-1}
\end{equation}
where

\begin{equation}
N_{0}=\sqrt{1+|\chi_{1}|^{2}+|\chi_{2}|^{2}}\,,\label{eq:54}
\end{equation}
is a normalization factor. The amplitudes to find an atom in the coupled
and uncoupled states read

\begin{align}
\psi_{\mathrm{C}_{j}} & =\langle\mathrm{C}_{j}|\Psi\rangle=\chi_{j}\psi_{0}+\psi_{j}\,,\label{eq:55}\\
\psi_{\mathrm{U}} & =\langle\mathrm{U}|\Psi\rangle=\frac{1}{N_{0}}(\psi_{0}-\chi_{1}^{*}\psi_{1}-\chi_{2}^{*}\psi_{2})\,.\label{eq:56}
\end{align}

The initial atomic amplitudes then can be expressed as

\begin{align}
\psi_{0} & =\frac{1}{N_{0}}\left(\psi_{\mathrm{U}}+\frac{1}{N_{0}}\chi_{1}^{*}\psi_{\mathrm{C}_{1}}+\frac{1}{N_{0}}\chi_{2}^{*}\psi_{\mathrm{C}_{2}}\right)\,,\label{eq:57}\\
\psi_{1} & =\psi_{\mathrm{C}_{1}}-\chi_{1}\psi_{0}=-\frac{\chi_{1}}{N_{0}}\psi_{\mathrm{U}}+\cdots\,,\label{eq:1-U}\\
\psi_{2} & =\psi_{\mathrm{C}_{2}}-\chi_{2}\psi_{0}=-\frac{\chi_{2}}{N_{0}}\psi_{\mathrm{U}}+\cdots\,.\label{eq:2-U}
\end{align}

The terms that are ommited in Eqs.~(\ref{eq:1-U}), (\ref{eq:2-U})
contain amplitudes $\psi_{\mathrm{C}_{1}}$, $\psi_{\mathrm{C}_{2}}$
and are not needed for the first order of the adiabatic approximation.

The equations for the atomic amplitudes in terms of coupled and uncoupled
states take the form

\begin{align}
i\partial_{t}\psi_{\mathrm{U}} & =\Delta\psi_{\mathrm{U}}+\cdots\,,\label{eq:U-1}\\
i\partial_{t}\psi_{\mathrm{C}_{1}} & =V_{1}\psi_{\mathrm{U}}-\frac{1}{2}\Omega_{1,1}^{*}N_{1}^{2}\psi_{e_{1}}-\frac{1}{2}\Omega_{2,0}^{*}\chi_{1}\psi_{e_{2}}+\cdots\,,\label{eq:C1}\\
i\partial_{t}\psi_{\mathrm{C}_{2}} & =V_{2}\psi_{\mathrm{U}}-\frac{1}{2}\Omega_{1,0}^{*}\chi_{2}\psi_{e_{1}}-\frac{1}{2}\Omega_{2,2}^{*}N_{2}^{2}\psi_{e_{2}}+\cdots\,,\label{eq:C2}\\
i\partial_{t}\psi_{e_{1}} & =-\frac{i}{2}\Gamma\psi_{e_{1}}-\frac{1}{2}\Omega_{1,1}\psi_{\mathrm{C}_{1}}\,,\label{eq:e1-C1}\\
i\partial_{t}\psi_{e_{2}} & =-\frac{i}{2}\Gamma\psi_{e_{2}}-\frac{1}{2}\Omega_{2,2}\psi_{\mathrm{C}_{2}}\,.\label{eq:e2-C2}
\end{align}

The ommited terms contain amplitudes $\psi_{\mathrm{C}_{1}}$, $\psi_{\mathrm{C}_{2}}$.
Here 

\begin{align}
V_{1} & =\frac{1}{N_{0}}(i\partial_{t}-\delta_{1})\chi_{1}\,,\label{eq:65}\\
V_{2} & =\frac{1}{N_{0}}(i\partial_{t}-\delta_{2})\chi_{2},\label{eq:66}
\end{align}
are the coefficients describing the non-adiabatic coupling; the detuning
$\Delta$ is defined as

\begin{equation}
\Delta=\frac{1}{N_{0}}i\partial_{t}\frac{1}{N_{0}}+\frac{\chi_{1}}{N_{0}}i\partial_{t}\frac{\chi_{1}^{*}}{N_{0}}+\frac{\chi_{2}}{N_{0}}i\partial_{t}\frac{\chi_{2}^{*}}{N_{0}}+\delta_{1}\frac{|\chi_{1}|^{2}}{N_{0}^{2}}+\delta_{2}\frac{|\chi_{2}|^{2}}{N_{0}^{2}}\,.\label{eq:67}
\end{equation}

Finally,

\begin{align}
N_{1} & =\sqrt{1+|\chi_{1}|^{2}}\,,\label{eq:68}\\
N_{2} & =\sqrt{1+|\chi_{2}|^{2}},\label{eq:69}
\end{align}
are normalization factors for the unnormalized coupled states.

\subsection{Adiabatic approximation}

Let us consider a situation where the light fields are slowly changing
and have large intensities. In this situation we can apply adiabatic
approximation. Neglecting small amplitudes of the coupled states $\psi_{\mathrm{C}_{1}}$
and $\psi_{\mathrm{C}_{2}}$ in Eqs.~(\ref{eq:C1}), (\ref{eq:C2})
leads to a pair of equations 

\begin{align}
2V_{1}\psi_{\mathrm{U}} & =\Omega_{1,1}^{*}N_{1}^{2}\psi_{e_{1}}+\Omega_{2,0}^{*}\chi_{1}\psi_{e_{2}}\,,\label{eq:70}\\
2V_{2}\psi_{\mathrm{U}} & =\Omega_{1,0}^{*}\chi_{2}\psi_{e_{1}}+\Omega_{2,2}^{*}N_{2}^{2}\psi_{e_{2}},\label{eq:71}
\end{align}
with the solution

\begin{align}
\psi_{e_{1}} & =2W_{1}\psi_{\mathrm{U}}\,,\label{eq:e1-U}\\
\psi_{e_{2}} & =2W_{2}\psi_{\mathrm{U}}\,,\label{eq:e2-U}
\end{align}
where

\begin{align}
W_{1} & =\frac{1}{\Omega_{1,1}^{*}N_{0}^{2}}(N_{2}^{2}V_{1}-\chi_{1}\chi_{2}^{*}V_{2})=\frac{1}{\Omega_{1,1}^{*}N_{0}^{3}}[N_{2}^{2}(i\partial_{t}-\delta_{1})\chi_{1}-\chi_{1}\chi_{2}^{*}(i\partial_{t}-\delta_{2})\chi_{2}]\,,\label{eq:W1}\\
W_{2} & =\frac{1}{\Omega_{2,2}^{*}N_{0}^{2}}(N_{1}^{2}V_{2}-\chi_{2}\chi_{1}^{*}V_{1})=\frac{1}{\Omega_{2,2}^{*}N_{0}^{3}}[N_{1}^{2}(i\partial_{t}-\delta_{1})\chi_{2}-\chi_{2}\chi_{1}^{*}(i\partial_{t}-\delta_{2})\chi_{1}]\,.\label{eq:W2}
\end{align}

On the other hand, Eqs.~(\ref{eq:e1-C1}), (\ref{eq:e2-C2}) relate
$\psi_{\mathrm{C}_{1}}$, $\psi_{\mathrm{C}_{2}}$ to $\psi_{e_{1}}$,
$\psi_{e_{2}}$ as

\begin{align}
\psi_{\mathrm{C}_{1}} & =-i\frac{\Gamma}{\Omega_{1,1}}\psi_{e_{1}}\,,\label{eq:76}\\
\psi_{\mathrm{C}_{2}} & =-i\frac{\Gamma}{\Omega_{2,2}}\psi_{e_{2}}\,.\label{eq:77}
\end{align}

Neglecting the decay of adiabatons, Eq.~(\ref{eq:U-1}) for the amplitude
of the uncoupled state becomes

\begin{equation}
i\partial_{t}\psi_{\mathrm{U}}=\Delta\psi_{\mathrm{U}}\,.\label{eq:main-U-1}
\end{equation}

Inserting Eqs.~(\ref{eq:e1-U}), (\ref{eq:e2-U}) into Eqs.~(\ref{eq:el10})-(\ref{eq:el22})
we get

\begin{align}
(\partial_{t}+c\partial_{z})\Omega_{1,0} & =i\frac{g}{N_{0}}W_{1}\,,\label{eq:el10a}\\
(\partial_{t}+c\partial_{z})\Omega_{2,0} & =i\frac{g}{N_{0}}W_{2}\,,\label{eq:80}\\
(\partial_{t}+c\partial_{z})\Omega_{1,1} & =-i\frac{g}{N_{0}}\chi_{1}^{*}W_{1}\,,\label{eq:81}\\
(\partial_{t}+c\partial_{z})\Omega_{2,2} & =-i\frac{g}{N_{0}}\chi_{2}^{*}W_{2}\,.\label{eq:el22a}
\end{align}

Equations (\ref{eq:main-U-1})-(\ref{eq:el22a}) describe adiabatic
propagation of the fields.

\subsection{Matrix form of equations}

From Eqs.~(\ref{eq:el10a})-(\ref{eq:el22a}) we obtain that the
total Rabi frequencies

\begin{align}
\Omega_{1} & =\sqrt{|\Omega_{1,1}|^{2}+|\Omega_{1,0}|^{2}}=|\Omega_{1,1}|N_{1}\,,\label{eq:83}\\
\Omega_{2} & =\sqrt{|\Omega_{2,2}|^{2}+|\Omega_{2,0}|^{2}}=|\Omega_{2,2}|N_{2},\label{eq:84}
\end{align}
obey the equations

\begin{align}
(\partial_{t}+c\partial_{z})\Omega_{1} & =0\,,\label{eq:85}\\
(\partial_{t}+c\partial_{z})\Omega_{2} & =0.\label{eq:86}
\end{align}

Combining Eqs.~(\ref{eq:el10a})-(\ref{eq:el22a}) we get the equations
for the ratios $\chi_{1}$ and $\chi_{2}$:

\begin{align}
(\partial_{t}+c\partial_{z})\chi_{1} & =ig\frac{N_{1}^{2}}{N_{0}}\frac{W_{1}}{\Omega_{1,1}}\,,\label{eq:chi1}\\
(\partial_{t}+c\partial_{z})\chi_{2} & =ig\frac{N_{2}^{2}}{N_{0}}\frac{W_{2}}{\Omega_{2,2}}.\label{eq:chi2}
\end{align}

Introducing the two-component spinor

\begin{equation}
\chi=\left(\begin{array}{c}
\chi_{1}\\
\chi_{2}
\end{array}\right),\label{eq:89}
\end{equation}
and using Eqs.~(\ref{eq:W1}), (\ref{eq:W2}), Eqs.~(\ref{eq:chi1}),
(\ref{eq:chi2}) can be written in the following matrix form:

\begin{equation}
(c^{-1}+\hat{v}^{-1})\partial_{t}\chi+\partial_{z}\chi+i\hat{v}^{-1}\hat{\delta}\chi=0\,,\label{eq:90}
\end{equation}
where

\begin{equation}
\hat{v}^{-1}=\frac{g}{cN_{0}^{4}}\left(\begin{array}{cc}
\frac{N_{1}^{2}N_{2}^{2}}{|\Omega_{1,1}|^{2}} & -\frac{N_{1}^{2}\chi_{1}\chi_{2}^{*}}{|\Omega_{1,1}|^{2}}\\
-\frac{N_{2}^{2}\chi_{2}\chi_{1}^{*}}{|\Omega_{2,2}|^{2}} & \frac{N_{2}^{2}N_{1}^{2}}{|\Omega_{2,2}|^{2}}
\end{array}\right)=\frac{g}{c}\left(\begin{array}{cc}
\frac{1}{\Omega_{1}^{2}}\frac{N_{1}^{4}}{N_{0}^{4}} & 0\\
0 & \frac{1}{\Omega_{2}^{2}}\frac{N_{2}^{4}}{N_{0}^{4}}
\end{array}\right)(N_{0}^{2}I-\chi\chi^{\dagger}),\label{eq:91}
\end{equation}
is the matrix of inverse group velocity,

\begin{equation}
\hat{\delta}=\left(\begin{array}{cc}
\delta_{1} & 0\\
0 & \delta_{2}
\end{array}\right),\label{eq:92}
\end{equation}
is the matrix of two-photon detunings.

\subsection{Equal fields}

Let us consider a particular case where $\Omega_{2,0}=\Omega_{1,0}$
and $\Omega_{2,2}=\Omega_{1,1}$ at the beginning of the medium ($z=0$).
We also assume that $\delta_{2}=\delta_{1}$. In this situation $\chi_{2}=\chi_{1}$
at all times and positions. Equation for the propagation becomes 
\begin{equation}
\left(c^{-1}+\frac{g}{c\Omega_{1}^{2}}\frac{N_{1}^{4}}{N_{0}^{4}}\right)\partial_{t}\chi_{1}+\partial_{z}\chi_{1}+i\frac{g}{c\Omega_{1}^{2}}\frac{N_{1}^{4}}{N_{0}^{4}}\delta_{1}\chi_{1}=0\,.\label{eq:93}
\end{equation}
 This equation has a similar form to the equation for the propagation
of the probe field in the EIT configuration. Also, it has the same
form as the equation for $\Lambda$-type system. The group velocity
is

\begin{equation}
v_{g}=\frac{c\Omega_{1}^{2}}{g}\frac{N_{0}^{4}}{N_{1}^{4}}\,.\label{eq:94}
\end{equation}

From this equation we see that the presence of the second pair of
beams increases the group velocity of the adiabaton propagation because
$N_{0}>N_{1}$. It can also be seen that the parts of the pulse with
larger amplitude propagate faster. This leads to changing shape of
the pulse and to an instability of a wave front. 

\begin{figure}
\includegraphics[width=0.7\columnwidth]{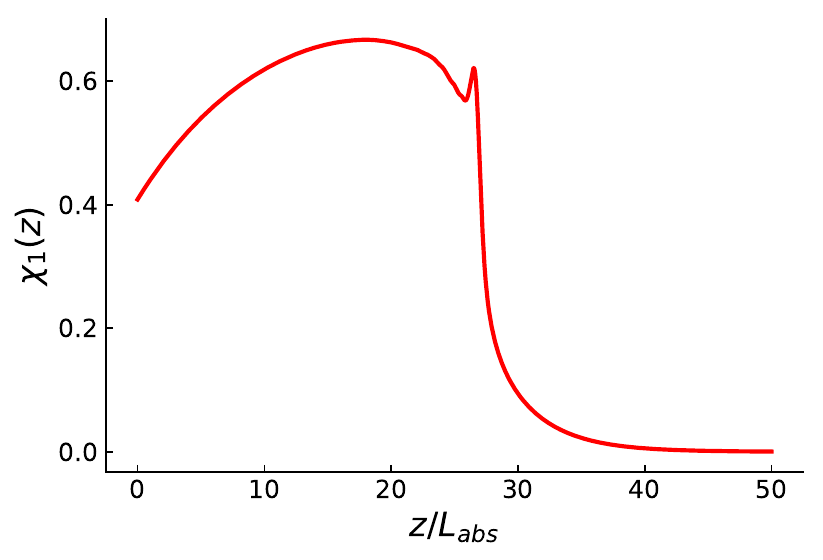}

\caption{Dependence of pulse amplitude ratio $\chi_{1}$ on the propagation
distance in M-type system calculated at time instance of $t=26.5\Gamma^{-1}$.}
\label{fig:steep}
\end{figure}

We conducted numerical study of the above field equations using the
Gaussian input fields similarly to the case of $\Lambda$ system:
the first field at the input of the atomic media is of the form $\Omega_{1,0}(0,t)=A\exp[-(t-t_{0})^{2}/\tau_{0}^{2}]$,
where $\tau_{0}=5\Gamma^{-1}$, $t_{0}=23\Gamma^{-1}$, $A=\Gamma$;
the second one is constant $\Omega_{1,1}(0,t)=1.5\Gamma$. Results
of the calculation are shown in Fig.~\ref{fig:steep} which demonstrates
significant pulse deformation and strong steepening effect at propagation
distances exceeding $25L_{\mathrm{abs}}$.

Deformation of pulse shape observed in M-type system has similar origin
as a self-steepening effect known in classical nonlinear optics \citep{1967-PR-DeMartini,2006-Suydam}.
The latter develops as a result of nonlinear dependence of the refractive
index on pulse intensity that leads to the intensity-dependent group
velocity. Commonly this effect is observed for the propagation of
short laser pulses producing sharp steepening of the trailing edge
of the pulse (optical shock) and accompanying spectral broadening.
The effect of self-steepening and accompanying pulse behavior has
been studied extensively in numerous papers \citep{1967-PR-DeMartini,2006-Suydam,1992-JOSAB-DeOliveira,1998-PRA-Trippenbach}.

\section{Adiabatons in double tripod system \label{sec:Double-tripod}}

\subsection{Initial equations}

\begin{figure}
\includegraphics[width=0.5\columnwidth]{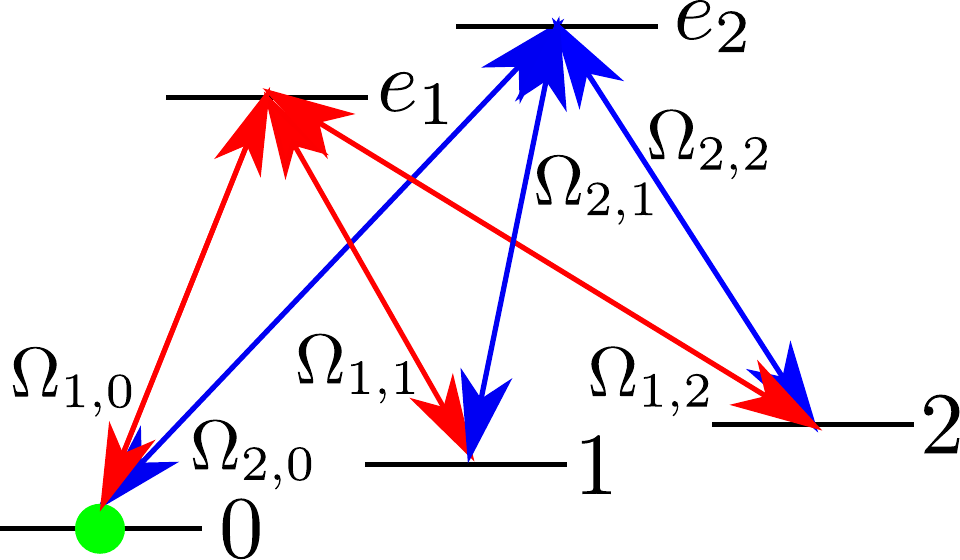}

\caption{Five level double tripod atomic system. Six laser beams with the Rabi
frequencies $\Omega_{1,0}$, $\Omega_{1,1}$, $\Omega_{1,2}$ and
$\Omega_{2,0}$, $\Omega_{2,1}$, $\Omega_{2,2}$ act on atoms characterized
by three hyperfine ground levels $0$, $1$ and $2$ as well as two
excited levels $e_{1}$ and $e_{2}$. Atoms are initially in the ground
level $0$.}
\label{fig:double-tripod}
\end{figure}
Finally, we investigate a double tripod type atomic system (Fig.~\ref{fig:double-tripod})
involving three metastable ground states $|0\rangle$, $|1\rangle$
and $|2\rangle$, as well as excited states $|e_{1}\rangle$ and $|e_{2}\rangle$.
Six laser fields with the Rabi frequencies $\Omega_{j,l}$ induce
resonant transitions $|l\rangle\rightarrow|e_{j}\rangle$. Applying
the rotating wave approximation (RWA), the atomic Hamiltonian in the
rotating frame with respect to the atomic levels reads

\begin{equation}
H_{\mathrm{2T}}=\sum_{j=1}^{2}\left\{ -\frac{1}{2}\left(\sum_{l=0}^{2}\Omega_{j,l}|e_{j}\rangle\langle l|+\mathrm{H.c.}\right)+\delta_{j}|j\rangle\langle j|-\frac{i}{2}\Gamma|e_{j}\rangle\langle e_{j}|\right\} \,,\label{eq:H-2}
\end{equation}
where $\delta_{j}$ are two-photon detunings. The losses in the Hamiltonian
(\ref{eq:H-2}) are taken into account in an effective way by introducing
a rate $\Gamma$ of the excited state decay. The time-dependent Schr\"odinger
equation $i\hbar\partial_{t}|\Psi\rangle=H_{\mathrm{2T}}|\Psi\rangle$
for the atomic state-vector $|\Psi\rangle=\sum_{l=0}^{2}\psi_{l}|l\rangle+\sum_{j=1}^{2}\psi_{e_{j}}|e_{j}\rangle$
yields the following equations for the atomic probability amplitudes
$\psi_{l}$, $\psi_{e_{j}}$:

\begin{align}
i\partial_{t}\psi_{l} & =\delta_{l}\psi_{l}-\frac{1}{2}\sum_{j}\Omega_{j,l}^{*}\psi_{e_{j}}\,,\label{eq:DT-l}\\
i\partial_{t}\psi_{e_{j}} & =-\frac{i}{2}\Gamma\psi_{e_{j}}-\frac{1}{2}\sum_{l}\Omega_{j,l}\psi_{l}\,.\label{eq:DT-e}
\end{align}

Here we introduced $\delta_{0}=0$. On the other hand, the Rabi frequencies
of the laser fields obey the following propagation equations

\begin{equation}
(\partial_{t}+c\partial_{z})\Omega_{j,l}=\frac{i}{2}g\psi_{e_{j}}\psi_{l}^{*}\,,\label{eq:el}
\end{equation}
where the parameter $g$ characterizes the strength of coupling of
the light fields with the atoms.

\subsection{Coupled and uncoupled states}

It is convenient to introduce the amplitudes of the two coupled states

\begin{equation}
\psi_{\mathrm{C}_{j}}=\frac{1}{\Omega_{j}}\sum_{l=0}^{2}\Omega_{j,l}\psi_{l}\,,\label{eq:99}
\end{equation}
where 
\begin{equation}
\Omega_{j}=\sqrt{\sum_{l=0}^{2}|\Omega_{j,l}|^{2}}\,.\label{eq:100}
\end{equation}

In addition, there is one uncoupled state

\begin{equation}
\psi_{U}=\frac{1}{N_{0}}\left|\begin{array}{ccc}
\psi_{0} & \psi_{1} & \psi_{2}\\
\Omega_{1,0}^{*} & \Omega_{1,1}^{*} & \Omega_{1,2}^{*}\\
\Omega_{2,0}^{*} & \Omega_{2,1}^{*} & \Omega_{2,2}^{*}
\end{array}\right|\equiv\sum_{l=0}^{2}A_{l}^{*}\psi_{l}\,,\label{eq:101}
\end{equation}
where $N_{0}$ ensures that the uncoupled state is normalized. Here

\begin{align}
A_{0} & =\frac{1}{N_{0}}(\Omega_{1,1}\Omega_{2,2}-\Omega_{1,2}\Omega_{2,1})\,,\label{eq:102}\\
A_{1} & =\frac{1}{N_{0}}(\Omega_{1,2}\Omega_{2,0}-\Omega_{1,0}\Omega_{2,2})\,,\label{eq:103}\\
A_{2} & =\frac{1}{N_{0}}(\Omega_{1,0}\Omega_{2,1}-\Omega_{1,1}\Omega_{2,0}),\label{eq:104}
\end{align}

and

\begin{equation}
N_{0}=\left[|\Omega_{1,1}\Omega_{2,2}-\Omega_{1,2}\Omega_{2,1}|^{2}+|\Omega_{1,2}\Omega_{2,0}-\Omega_{1,0}\Omega_{2,2}|^{2}+|\Omega_{1,0}\Omega_{2,1}-\Omega_{1,1}\Omega_{2,0}|^{2}\right]^{\frac{1}{2}}\,.\label{eq:105}
\end{equation}

Note, that 
\begin{equation}
N_{0}^{2}=\Omega_{1}^{2}\Omega_{2}^{2}-\left(\sum_{l=0}^{2}\Omega_{2,l}^{*}\Omega_{1,l}\right)\left(\sum_{l=0}^{2}\Omega_{1,l}^{*}\Omega_{2,l}\right)\,.\label{eq:106}
\end{equation}
 The initial atomic amplitudes then can be expressed as

\begin{equation}
\psi_{l}=A_{l}\psi_{\mathrm{U}}+\cdots\,\label{eq:psi-l-psi-U}
\end{equation}

The ommited terms in Eq.~(\ref{eq:psi-l-psi-U}) contain amplitudes
$\psi_{\mathrm{C}_{1}}$, $\psi_{\mathrm{C}_{2}}$. Note, that the
coefficients $A_{l}$ have the property 

\begin{equation}
\sum_{l=0}^{2}A_{l}\Omega_{j,l}=0\,.\label{eq:108}
\end{equation}

The equations (\ref{eq:DT-l}), (\ref{eq:DT-e}) for the atomic amplitudes
in terms of coupled and uncoupled states take the form

\begin{align}
i\partial_{t}\psi_{\mathrm{U}} & =\Delta\psi_{\mathrm{U}}+\cdots\,,\label{eq:U-1-1}\\
i\partial_{t}\psi_{\mathrm{C}_{j}} & =V_{j}\psi_{\mathrm{U}}-\frac{1}{2\Omega_{j}}\sum_{m}\psi_{e_{m}}\sum_{l=0}^{2}\Omega_{j,l}\Omega_{m,l}^{*}\,,\label{eq:Cj}\\
i\partial_{t}\psi_{e_{j}} & =-\frac{i}{2}\Gamma\psi_{e_{j}}-\frac{1}{2}\Omega_{j}\psi_{\mathrm{C}_{j}}\,.\label{eq:ej-Cj}
\end{align}

Here,

\begin{align}
\Delta & =\sum_{l=0}^{2}A_{l}(i\partial_{t}+\delta_{l})A_{l}^{*}\,,\label{eq:112}\\
V_{j} & =\frac{1}{\Omega_{j}}\sum_{l=0}^{2}A_{l}(i\partial_{t}+\delta_{l})\Omega_{j,l}=\sum_{l=0}^{2}A_{l}(i\partial_{t}+\delta_{l})\frac{\Omega_{j,l}}{\Omega_{j}}\,.\label{eq:113}
\end{align}

The terms omitted in Eq.~(\ref{eq:U-1-1}) contain amplitudes $\psi_{\mathrm{C}_{1}}$
and $\psi_{\mathrm{C}_{2}}$.

\subsection{Adiabatic approximation}

Let us consider a situation where the light fields are slowly changing
and have large intensities. In this situation we can apply adiabatic
approximation. Neglecting small amplitudes of the coupled states $\psi_{\mathrm{C}_{1}}$
and $\psi_{\mathrm{C}_{2}}$ in Eq.~(\ref{eq:Cj}) leads to a system
of equations

\begin{equation}
2\Omega_{j}V_{j}\psi_{\mathrm{U}}=\sum_{m}\psi_{e_{m}}\sum_{l=0}^{2}\Omega_{j,l}\Omega_{m,l}^{*},\label{eq:114}
\end{equation}
with the solution 
\begin{equation}
\psi_{e_{j}}=2W_{j}\psi_{\mathrm{U}}\,,\label{eq:ej-U}
\end{equation}
 where

\begin{align}
W_{1} & =\frac{1}{N_{0}^{2}}(\Omega_{2}^{2}\Omega_{1}V_{1}-\sum_{l=0}^{2}\Omega_{2,l}^{*}\Omega_{1,l}\Omega_{2}V_{2})=\frac{1}{N_{0}^{2}}\sum_{l=0}^{2}A_{l}\left(\Omega_{2}^{2}(i\partial_{t}+\delta_{l})\Omega_{1,l}-\sum_{m=0}^{2}\Omega_{2,m}^{*}\Omega_{1,m}(i\partial_{t}+\delta_{l})\Omega_{2,l}\right)\,,\label{eq:116}\\
W_{2} & =\frac{1}{N_{0}^{2}}(\Omega_{1}^{2}\Omega_{2}V_{2}-\sum_{l=0}^{2}\Omega_{1,l}^{*}\Omega_{2,l}\Omega_{1}V_{1})=\frac{1}{N_{0}^{2}}\sum_{l=0}^{2}A_{l}\left(\Omega_{1}^{2}(i\partial_{t}+\delta_{l})\Omega_{2,l}-\sum_{m=0}^{2}\Omega_{1,m}^{*}\Omega_{2,m}(i\partial_{t}+\delta_{l})\Omega_{1,l}\right)\,.\label{eq:117}
\end{align}

On the other hand, Eq.~(\ref{eq:ej-Cj}) relates $\psi_{\mathrm{C}_{j}}$
to $\psi_{e_{j}}$ as

\begin{equation}
\psi_{\mathrm{C}_{j}}=-i\frac{\Gamma}{\Omega_{j}}\psi_{e_{j}}\,.\label{eq:118}
\end{equation}

Neglecting the decay of adiabatons, Eq.~(\ref{eq:U-1-1}) for the
amplitude of the uncoupled state becomes 
\begin{equation}
i\partial_{t}\psi_{\mathrm{U}}=\Delta\psi_{\mathrm{U}}\,.\label{eq:main-U-2}
\end{equation}

Inserting Eq.~(\ref{eq:ej-U}) into Eq.~(\ref{eq:el}) we get 
\begin{equation}
(\partial_{t}+c\partial_{z})\Omega_{j,l}=igW_{j}A_{l}^{*}\,.\label{eq:el-a}
\end{equation}

Equations (\ref{eq:main-U-2}) and (\ref{eq:el-a}) describe adiabatic
propagation of the fields. For the validity of the adiabatic apporximation
one should require that the population of the excited states be small
compared to the population of the uncoupled state. According to Eq.~(\ref{eq:U-1-1})
this leads to the condition 
\begin{equation}
|W_{j}|\ll1\,.\label{eq:DT-adiabat}
\end{equation}

From Eq.~(\ref{eq:el-a}) we obtain that the total Rabi frequencies
$\Omega_{j}$ obey the equations 
\begin{equation}
(\partial_{t}+c\partial_{z})\Omega_{j}=0.\label{eq:prop-total-1}
\end{equation}

In addition, we have 
\begin{equation}
(\partial_{t}+c\partial_{z})\sum_{l=0}^{2}\Omega_{2,l}^{*}\Omega_{1,l}=0.\label{eq:prop-total-2}
\end{equation}
 From Eqs.~(\ref{eq:prop-total-1}), (\ref{eq:prop-total-2}) it
follows 
\begin{equation}
(\partial_{t}+c\partial_{z})N_{0}=0\,.\label{eq:124}
\end{equation}

\subsection{Matrix form of equations}

Introducing the two-component spinors (columns)

\begin{equation}
\hat{\Omega}_{l}=\left(\begin{array}{c}
\Omega_{1,l}\\
\Omega_{2,l}
\end{array}\right).\label{eq:125}
\end{equation}

We can write Eqs.~(\ref{eq:el-a}) as

\begin{equation}
(c^{-1}\partial_{t}+\partial_{z})\hat{\Omega}_{l}+A_{l}^{*}\hat{v}^{-1}\sum_{n=0}^{2}A_{n}(\partial_{t}-i\delta_{n})\hat{\Omega}_{n}=0\,,\label{eq:eq:el-matrix}
\end{equation}
where

\begin{equation}
\hat{v}^{-1}=\frac{g}{c}\frac{1}{N_{0}^{2}}\left(\begin{array}{cc}
\Omega_{2}^{2} & -\sum_{m=0}^{2}\Omega_{2,m}^{*}\Omega_{1,m}\\
-\sum_{m=0}^{2}\Omega_{1,m}^{*}\Omega_{2,m} & \Omega_{1}^{2}
\end{array}\right),\label{eq:127}
\end{equation}
is the matrix of inverse group velocity \citep{2014-NC-Lee}. Note
that

\begin{equation}
\det\hat{v}^{-1}=\frac{g^{2}}{c^{2}N_{0}^{2}},\label{eq:128}
\end{equation}
and

\begin{equation}
(\partial_{t}+c\partial_{z})\hat{v}^{-1}=0\,.\label{eq:129}
\end{equation}

Multiplying Eq.~(\ref{eq:eq:el-matrix}) by $A_{l}$ and summing
we get

\begin{equation}
\sum_{l=0}^{2}A_{l}(c^{-1}\partial_{t}+\partial_{z})\hat{\Omega}_{l}+\hat{v}^{-1}\sum_{l=0}^{2}A_{l}(\partial_{t}-i\delta_{l})\hat{\Omega}_{l}=0\,.\label{eq:131}
\end{equation}

Finally, using the matrix of inverse group velocity $\hat{v}^{-1}$
the adiabatic condition (\ref{eq:DT-adiabat}) in double tripod system
can be written as

\begin{equation}
\frac{c}{g}\left|\hat{v}^{-1}\sum_{l=0}^{2}A_{l}(i\partial_{t}+\delta_{l})\hat{\Omega}_{l}\right|\ll1\,.\label{eq:DT-adiabat-2}
\end{equation}
The condition should hold for each component of the column in vertical
bars.

\subsection{Adiabatons}

Let us consider the situation where there are no two-photon detunings,
$\delta_{l}=0$. We will search for a propagating solution in the
form $\Omega_{j,l}=f_{j,l}\left(t-\frac{1}{v_{g}}z\right)$. Then
$\partial_{z}\Omega_{j,l}=-v_{g}^{-1}\partial_{t}\Omega_{j,l}$ and
Eq.~(\ref{eq:eq:el-matrix}) becomes

\begin{equation}
v_{g}^{-1}\partial_{t}\hat{\Omega}_{l}=A_{l}^{*}\hat{v}^{-1}\sum_{n=0}^{2}A_{n}\partial_{t}\hat{\Omega}_{n}\,.\label{eq:prop-1}
\end{equation}

Multiplying this equation by $A_{l}$ and summing over $l$ we obtain

\begin{equation}
\hat{v}^{-1}\sum_{l=0}^{2}A_{l}\partial_{t}\hat{\Omega}_{l}=v_{g}^{-1}\sum_{l=0}^{2}A_{l}\partial_{t}\hat{\Omega}_{l}\,.\label{eq:133}
\end{equation}
We see that in this case $\sum_{l=0}^{2}A_{l}\partial_{t}\hat{\Omega}_{l}$
should be an eigenvector of a matrix $\hat{v}^{-1}$, with $v_{g}$
being the eigenvalue. Note, that since the adiabatic condition (\ref{eq:DT-adiabat-2})
contains inverse group velocity, the adiabaticity requirement for
the solution with smaller value of $v_{g}$ is stricter.

Let us write the eigenvector of the matrix $\hat{v}^{-1}$ corresponding
to an eigenvalue $v_{g}^{-1}$ as $(1,\xi)^{T}$. Since this eigenvector
is equal to $\sum_{l=0}^{2}A_{l}\partial_{t}\hat{\Omega}_{l}$, we
have $\sum_{l=0}^{2}A_{l}\partial_{t}\Omega_{2,l}=\xi\sum_{l=0}^{2}A_{l}\partial_{t}\Omega_{1,l}$.
Then Eq.~(\ref{eq:prop-1}) can be written as

\begin{align}
\partial_{t}\Omega_{1,l} & =A_{l}^{*}\sum_{n=0}^{2}A_{n}\partial_{t}\Omega_{1,n}\,,\label{eq:134}\\
\partial_{t}\Omega_{2,l} & =\xi A_{l}^{*}\sum_{n=0}^{2}A_{n}\partial_{t}\Omega_{1,n},\label{eq:135}
\end{align}
or

\begin{equation}
\partial_{t}\Omega_{2,l}=\xi\partial_{t}\Omega_{1,l},\label{eq:136}
\end{equation}
with the solution 

\begin{equation}
\Omega_{2,l}(t)-\xi\Omega_{1,l}(t)=\Omega_{2,l}(0)-\xi\Omega_{1,l}(0)\,.\label{eq:137}
\end{equation}

Initially the pulses $\Omega_{1,0}$ and $\Omega_{2,0}$ are absent,
thus $\Omega_{2,l}(t)=\xi\Omega_{1,l}(t)$. 

As a consequence of Eqs.~(\ref{eq:prop-total-1}), (\ref{eq:prop-total-2})
the propagating solution obeys the equations 
\begin{equation}
\Omega_{j}=\mathrm{const}\,,\qquad\sum_{l=0}^{2}\Omega_{2,l}^{*}\Omega_{1,l}=\mathrm{const}\,.\label{eq:138}
\end{equation}
 These equations ensure that the velocity $v_{g}$ remains constant.

Let us consider the situation when $\Omega_{2,2}(0)=\Omega_{1,1}(0)$
and $\Omega_{2,1}(0)=\Omega_{1,2}(0)$. Then the eigenvalues are $v_{g}=\frac{c}{g}[\Omega_{1,1}(0)-\xi\Omega_{1,2}(0)]^{2}$
with $\xi=\pm1$ . From the equations it follows that $\Omega_{2,2}=\Omega_{1,1}$
and $\Omega_{2,1}=\Omega_{1,2}$ for all times. The values of $\Omega_{1,1}$
and $\Omega_{1,2}$ can be found by solving the equations $\Omega_{1,2}-\xi\Omega_{1,1}=\Omega_{1,2}(0)-\xi\Omega_{1,1}(0)$
, $\Omega_{1,0}^{2}+\Omega_{1,1}^{2}+\Omega_{1,2}^{2}=\Omega_{1,1}^{2}(0)+\Omega_{1,2}^{2}(0)$.

\begin{figure}
\includegraphics[width=0.7\columnwidth]{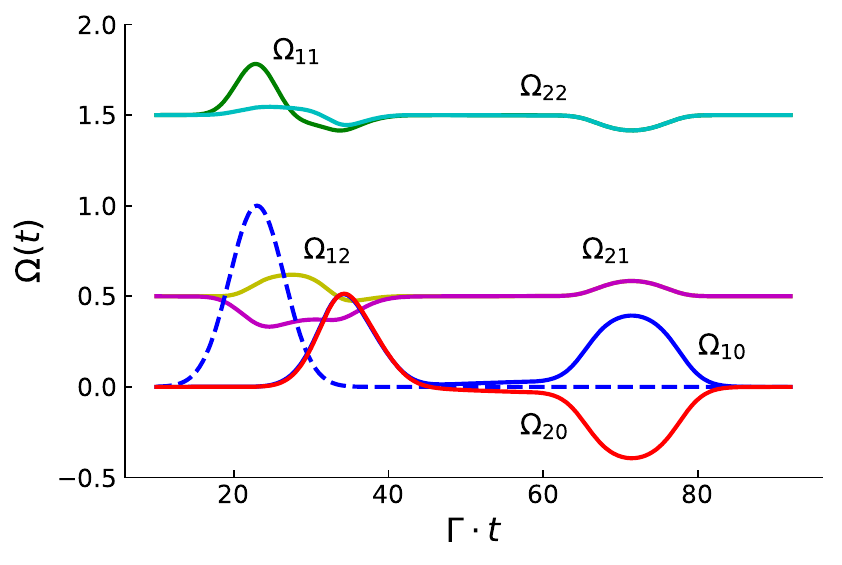}

\caption{Temporal dependence of pulse amplitudes in double tripod system. Dashed
line corresponds to pulse envelope $\Omega_{1,0}$ at the input ($z=0$),
solid lines show pulse envelopes at propagation distance $z=50L_{\mathrm{abs}}$.
Amplitude shown on the vertical scale is measured in $\Gamma$.}
\label{fig:DT-adiabaton}
\end{figure}

Similarly to the previous cases, we used numerical calculations to
demonstrate the obtained effects. At the begining of the atomi media
two optical fields have Gaussian temporal shape: $\Omega_{1,0}(0,t)=A_{1}\exp[-(t-t_{1})^{2}/\tau_{1}^{2}]$,
$\Omega_{2,0}(0,t)=A_{2}\exp[-(t-t_{2})^{2}/\tau_{2}^{2}]$, where
$\tau_{1}=\tau_{2}=5\Gamma^{-1}$, $t_{1}=t_{2}=23\Gamma^{-1}$, $A_{1}=1$,
$A_{2}=0$; other fields are constant with amplitudes $\Omega_{1,1}(0,t)=\Omega_{2,2}(0,t)=1.5\Gamma$
and $\Omega_{1,2}(0,t)=\Omega_{2,1}(0,t)=0.5\Gamma$. Corresponding
numerical results are depicted in Fig.~\ref{fig:DT-adiabaton}. It
can be seen that there is adiabatonic regime for the fields $\Omega_{j,l}$
after some propagation distance. There are two combinations of fields
propagating with different group velocities. The first combination
seen on the right part of Fig.~\ref{fig:DT-adiabaton} has $\Omega_{2,0}(z,t)=-\Omega_{1,0}(z,t)$
and is propagating with smaller group velocity. The second combination
having $\Omega_{2,0}(z,t)=\Omega_{1,0}(z,t)$ is propagating with
larger group velocity and can be seen in the left part of Fig.~\ref{fig:DT-adiabaton}.

\section{Discussion, outlook and conclusions \label{sec:Conclusions}}

In summary, we have analyzed the propagation of optical pulses in
multilevel systems of M-type and double tripod under the adiabatic
approximation. Both systems contain five non-degenerate atomic levels
and exhibit an uncoupled (dark) state. We also compared our results
to a three level $\Lambda$ system which has been used as a well known
reference system for our theoretical approach. It was found that in
the case of M-type atomic system the group velocity depends on the
amplitudes of the fields, thus different part of the pulse propagate
with different velocity. As a result, a region of the pulse with large
steepness appears and the adiabatic approximation becomes not valid
after some distance. In contrast, in the double-tripod system there
are two different configurations of the fields that propagate without
changing shape and the adiabatic propagation of light pulses is possible.
The latter system can be realized experimentally like the one used
as a setup demonstrating SSL formation in the atomic system \citep{2014-NC-Lee}.
In that study a double tripod scheme utilizing laser-cooled Rb atoms
loaded into magneto-optical trap was explored. Using similar setup
with the appropriate parameters the results of our theoretical model
presented in Fig.~\ref{fig:DT-adiabaton} can be verified experimentally.
To conclude, our study demonstrates that in the application to many
level systems the adiabatic propagation of light pulses is configuration-sensitive
and is not common to all multi-level systems having an uncoupled (dark)
state. Future work could extend this analysis by exploring adiabatons that carry orbital angular momentum (OAM) through the inclusion of paraxial propagation equations with transverse dynamics, beyond the current one-dimensional approach. This extension would allow us to investigate the impact of transverse spatial structure on adiabatic propagation, opening new possibilities for controlling light-matter interactions in complex atomic systems.
\section*{Acknowledgments}
This project has received funding from the Research Council of Lithuania (LMTLT), agreement No. S-ITP-24-6.
\\
\section*{Disclosures}
The authors declare no conflicts of interest.

\end{document}